\documentclass[reprint,
superscriptaddress,
 amsmath,amssymb,
 aps,
pra,
floatfix,
]{revtex4-1}
\usepackage{float}
\usepackage{graphicx}
\graphicspath{ {images/}, {figs/} }
\usepackage{dcolumn}
\usepackage{bm}
\usepackage{hyperref}
\usepackage[mathlines]{lineno}
\usepackage{amssymb}
\usepackage{xcolor}
\hypersetup{
	colorlinks,
	linkcolor={red!50!black},
	citecolor={green!50!black},
	urlcolor={blue!80!black}
}
\usepackage[caption=false]{subfig}
\makeatletter
\newcommand{\colorcaption}[2][]{%
  \begingroup%
  \renewcommand{\@caption@fignum@sep}{ (color online). }%
  \caption[#1]{#2}%
  \endgroup%
}
\makeatother

\begin{document}

\title{Topological corner modes in a brick lattice with nonsymmorphic symmetry}
\author{Yuhan Liu}
\affiliation{Department of Physics, Sun Yat-sen University,  Guangzhou 510275, China}
\affiliation{Department of Physics, the University of Chicago, Chicago, Illinois 60637, USA}

\author{Yuzhu Wang}
\affiliation{Department of Physics, Sun Yat-sen University,  Guangzhou 510275, China}

\author{Nai Chao Hu}
\affiliation{Department of Physics, Sun Yat-sen University,  Guangzhou 510275, China}
 \affiliation{Department of Physics, University of Texas at Austin, Austin, TX 78712, USA}

\author{Jun Yu Lin}
\affiliation{Department of Physics, Sun Yat-sen University,  Guangzhou 510275, China}
\affiliation{Department of Physics, The Chinese University of Hong Kong, Hong Kong, China}

\author{Ching Hua Lee}
\email{phylch@nus.edu.sg}
\affiliation{Institute of High Performance Computing, 138632, Singapore}
\affiliation{Department of Physics, National University of Singapore, Singapore, 117542.}
\author{Xiao Zhang}
\email{zhangxiao@mail.sysu.edu.cn}
\affiliation{Department of Physics, Sun Yat-sen University,  Guangzhou 510275, China}

\date{\today}{

\begin{abstract}
The quest for new realizations of higher-order topological system has garnered much recent attention. In this work, we propose a paradigmatic brick lattice model where corner modes requires protection by nonsymmorphic symmetry in addition to two commuting mirror symmetries. Unlike the well-known square corner mode lattice, it has an odd number of occupied bands, which necessitates a different definition for the $\mathbb Z_2\times \mathbb Z_2$ topological invariant. By studying both the quadrupolar polarization and effective edge model, our study culminates in a phase diagram containing two distinct topological regimes. Our brick lattice corner modes can be realized in a RLC circuit setup and detected via collossal ``topolectrical" resonances.
\end{abstract}

\maketitle


\section{\label{sec:intro}Introduction}

In much of topological condensed matter systems from Quantum Hall gases\cite{thouless1982quantized,streda1982theory,klitzing1980new} to topological insulators\cite{kane2005z,bernevig2006quantum,fu2007topological,moore2007topological,konig2007quantum,qi2008topological,moore2010birth,hasan2010colloquium,qi2011topological} and Weyl semimetals\cite{burkov2011weyl,wan2011topological,balents2011weyl,burkov2011weyl,turner2013beyond,vafek2014dirac,weng2015weyl}, the focus has been on protected modes at the boundary of a topological bulk. Such modes exist by virtue of nontrivial Wannier polarization, analogous to boundary charge accumulation from classical electric dipole polarization. But recently, this analogy has been further extended to quadrupolar or higher polarizations, where the intrinsic directionality of a multipole gives rise to enigmatic topological phenomena occurring only when \emph{two or more} open boundaries are present\cite{Benalcazar2017Quantized}. In such systems, topologically protected ``higher-order'' corner modes can exist at the intersection of edges, even if the edges themselves do not host topological modes\cite{langbehn2017reflection,noh2018topological,imhof2018topolectrical,schindler2018higher,lin2018topological,schindler2018higherti}.

From a complementary viewpoint, these corner modes can also be inferred from special crystal symmetries, with their host lattices regarded as glorified topological crystalline insulators (TCIs)\cite{fu2011topological,tanaka2012experimental,dziawa2012topological,hsieh2012topological,ando2015topological,zhang2016invited,bismuth,Schindlereaat0346}. In the archetypal higher-order square lattice\cite{Benalcazar2017Quantized}, the corner mode is protected by two non-commutable mirror symmetries that defines a nontrivial mirror Chern number. As a slightly more sophisticated example, corner modes also exist in the breathing Kagome lattice\cite{ezawa2017higher}, where they are protected by three mirror symmetries. An advantage of viewing higher-order phenomenon as symmetry-protected topological order is that it does not presuppose the existence of a Fermi sea, unlike the viewpoint of nested Wannier polarization. As such, bona fide higher-order topological corner modes should exist in classical and quantum lattices alike, even when higher-order polarization do not correspond to any physical charge accumulation. Indeed, topological corner modes have been experimentally observed with relative ease in various classical photonic, mechanical and electrical lattices~\cite{peterson2018quantized,Serra2018Observation,imhof2018topolectrical}, where couplings can be fine-tuned with precision.


Encouraged by these practical advances, we propose in this work a higher-order topological brick lattice with novel nonsymmorphic symmetry in addition to two commuting mirror symmetries~\cite{PhysRevLett.119.246401}, unlike the often used square corner mode lattice which possesses $C_4$ rotational symmetry and two non-commuting mirror symmetries. More fundamentally, it has an odd instead of even number of occupied bands at half filling, which necessitates an alternative definition of its $\mathbb{Z}_2\times\mathbb{Z}_2$ topological index distinct from well-studied models\cite{Benalcazar2017Quantized,benalcazar2017electric}. First, we begin by describing our brick lattice and providing numerical evidence for higher-order corner modes. Following that, we justify their robustness both in terms of a newly defined $\mathbb{Z}_2\times\mathbb{Z}_2$ topological index and an edge Hamiltonian picture, with three distinct gapped phases illustrated in a phase diagram. Next, we discuss the consequences of breaking non-symmorphic symmetry before finally proposing an experimental setup for detecting these brick lattice corner modes with circuit impedance measurements.


\section{\label{sec:tb} Brick lattice model and corner modes}
\subsection{Brick lattice structure and tight-binding Hamiltonian}

We study a brick lattice as shown in Fig.~\ref{lattice}. The six sites in each unit cell are connected via various real hoppings as described in Fig.~\ref{lattice}b. Notice that are two inequivalent types of ``bricks'', one which is wholly contained within a unit cell, and the other which straddles three unit cells and contains a possibly nonvanishing coupling $t_3$ through its width. Note that all couplings are meant to be properties of the lattice structure, and are unaffected by the lattice distortion angle $\theta$. For this reason, our brick lattice is suitable for circuit implementation, as described later. In general, such geometry agnostic property is useful for lattice model engineering, where desired properties can be designed through universal complex analytic properties that are embedded in the graph structure~\cite{budich2014search,lee2016band,lee2017band,kunst2018lattice}, not geometric structure of the lattice.

As we can see in Fig.~\ref{lattice}a, the brick lattice possesses two commuting mirror symmetries $M_x$ and $M_y$ about the x and y-axes, as well as the nonsymmorphic (glide reflection) symmetry $g_y=\{m_y|\tau_x\}$. Specifically, the lattice is mapped onto itself when translated along half a unit cell ($\tau_x$, red dashed arrow) and then reflected along the mirror plane ($m_y$, blue dashed line). When $t_1=a$ and $t_3=0$, our brick lattice possesses the same $C_6$ rotational symmetry as the corner mode lattice of~\cite{noh2018topological}; but as we shall show, the corner mode behavior can persist far beyond this limit, and hence does not require $C_6$ rotational symmetry at all. Indeed, nonsymmorphic symmetry has been known to protect various interesting topological features from tilted Dirac cones to surface states with M\"{o}bius twists~\cite{parameswaran2013topological,liu2014topological,shiozaki2015z,watanabe2015filling,schoop2016dirac,lin2017line}. 

\begin{figure}
\centering
\includegraphics[width=\linewidth]{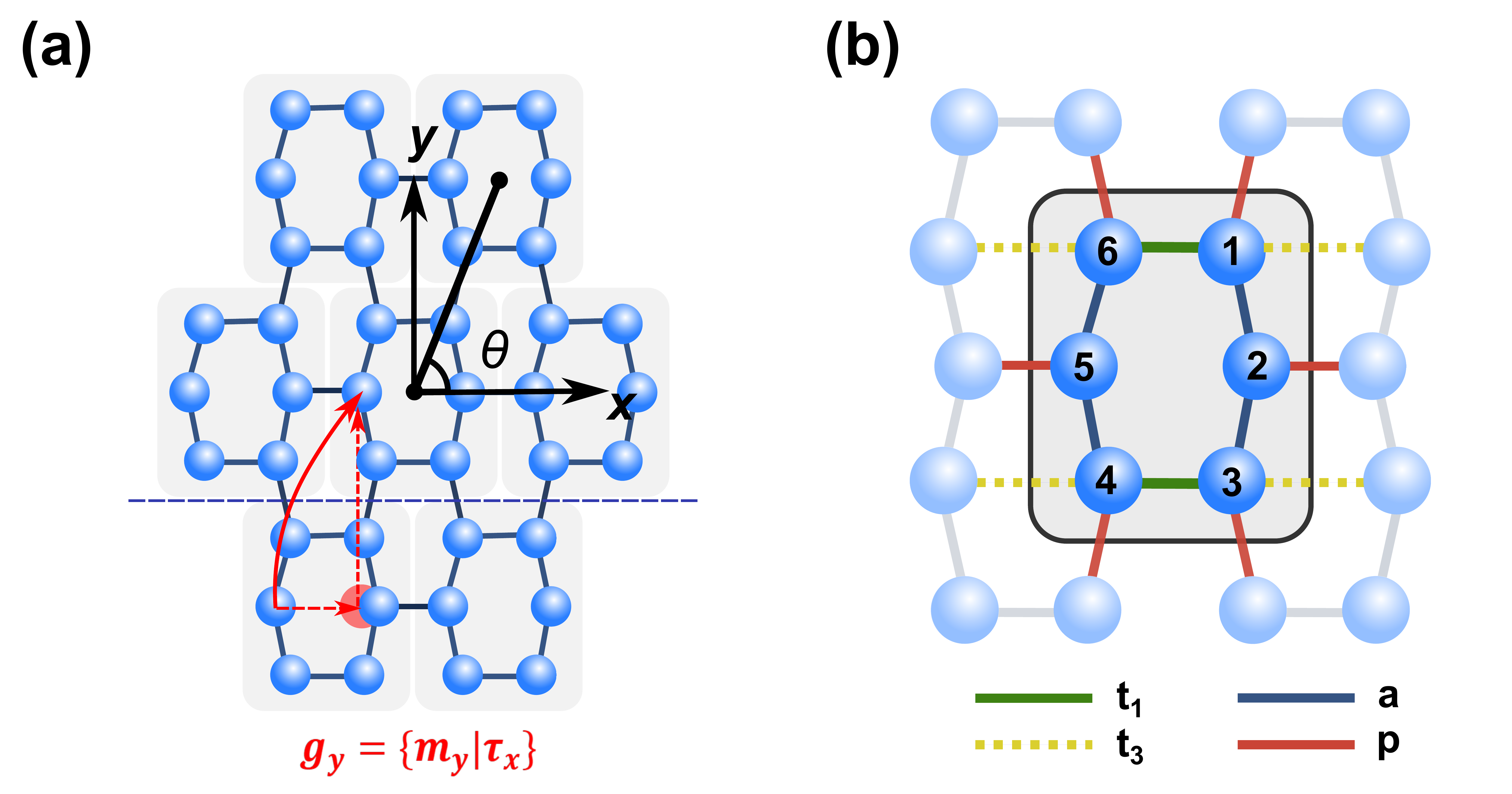}
\caption{a) General structure of our brick lattice, which possess commuting mirror reflecting symmetries $M_x$ and $M_y$ about the x and y-axes, as well as a nonsymmorphic  symmetry $g_y=\{m_y|\tau_x\}$ (red arrow) consisting of a glide along half a unit cell (horizontal left arrow) and a reflection (vertical left arrow). Note that reflection plane for $g_y$ is the blue dashed line, not the $x$-axis. b) The sublattice basis and couplings $t_1,a,t_3,p$ defining the brick Hamiltonian Eq.~\ref{Hamiltonian}. There are two types of ``bricks'', one contained with a single unit cell (shown), and the other straddling three neighboring unit cells, with an additional coupling $t_3$ across its width. $t_3$ and $p$ always connect sites of adjacent unit cells.}
\label{lattice}
\end{figure}

In the basis of sublattices 1 to 6 illustrated Fig.~\ref{lattice}b, the couplings are contained in an effective Hamiltonian
\begin{widetext}
\begin{equation}
H(k_1,k_2)=\left(
\begin{array}{cccccc}
      e & a & 0 & p\, e^{i k_1}& 0 & t_1+t_3 e^{i(k_1+k_2)}\\
      a & f & a & 0 & p\, e^{i(k_1+k_2)} &0\\
     0 & a & e & t_1+t_3 e^{i(k_1+k_2)} & 0 & p\, e^{ik_2}\\
     p\, e^{-ik_1} & 0 & t_1+t_3 e^{-i(k_1+k_2)} & e & a & 0\\
     0 & p\,  e^{-i(k_1+k_2)} & 0 & a & f & a\\
     t_1+t_3 e^{-i(k1+k2)} & 0 & p\, e^{-ik_2}& 0 & a & e
\end{array}
\right),
\label{Hamiltonian}
\end{equation}
\end{widetext}
with onsite energies $e$ and $f$ at the corners and midpoints of each brick respectively. $k_1,k_2$ are related to the lattice momenta $k_x,k_y$ via $k_1=k_x\cos{\theta}+k_y\sin{\theta}, k_2=k_x\cos{\theta}-k_y\sin{\theta}$, such that $\theta$ indeed never appears explicitly. Since higher-order topological phenomena are essentially mathematical properties of the lattice rather than that of the particles inhabiting it, our following results will be equally valid even if $H(k_1,k_2)$ is interpreted as a lattice Laplacian or any other linear operator on the lattice graph.

\subsection{Band structure and corner modes}

We next sequentially present the band structure and eigenmodes of our brick Hamiltonian under periodic, single and double open boundary conditions (PBCs, single and double OBCs), so as to elucidate how exactly the corner modes emerge.

To present various possible contrasting scenarios, we shall consider three sets of parameters, as illustrated in the top row of Fig.~\ref{edge}:
\begin{itemize}
\item Case A:  $p=-5.0,t_1=-1.0,t_3=-0.1,a=-1.0$
\item Case B:  $p=-0.5,t_1=-1.0,t_3=-0.1,a=-1.0$
\item Case C:  $p=-4.0,t_1=-1.0,t_3=-7.0,a=-1.0$
\begin{equation}\label{sets}\end{equation}
\end{itemize}
Case A contains much stronger couplings across unit cells than case B. Case C is somewhat similar to case A, but with much stronger $t_3$-type couplings across the widths of bricks that straddle unit cells. Henceforth, we shall also set the onsite energies $e$ and $f$ to zero, so that the corner modes can be pinned at zero energy ($\omega=0$).

First, we examine the bulk (PBC) band structure of the brick lattice. In all three cases, a gap clearly separates the upper three bands from the lower three bands (Fig.~\ref{edge} middle row), allowing unambiguous topological characterization of potential midgap modes. 
\begin{figure*}
\centering
\includegraphics[width=\linewidth]{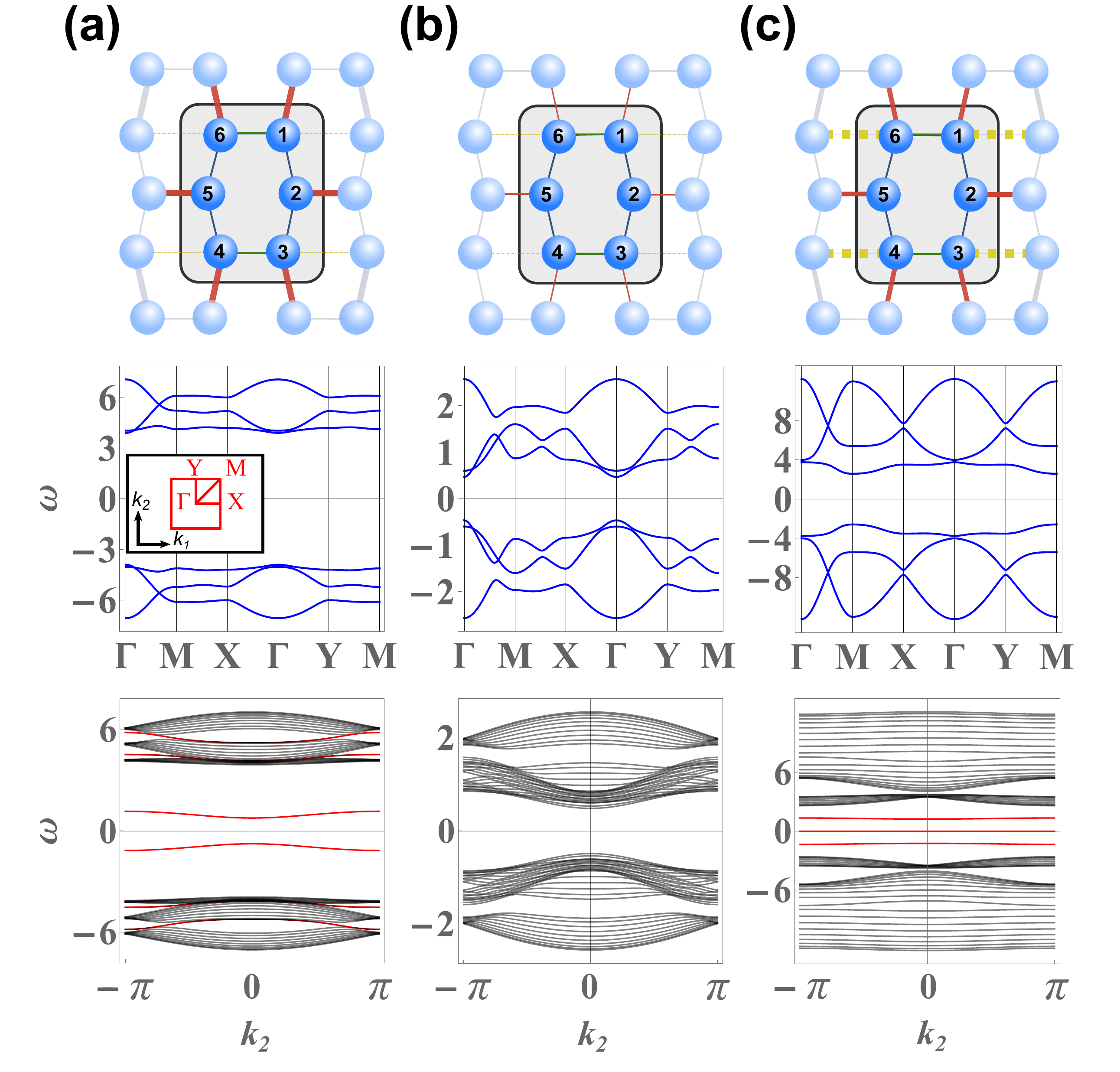}
\caption{Bulk and edge spectra of our brick lattice Hamiltonian, with columns a) to c) corresponding to parameters given by cases A to C (\ref{sets}). Top Row) Lattice couplings for each case, colored according to Fig.~\ref{lattice}b with thicknesses proportional to coupling magnitude. Middle Row) PBC bandstructures with well-defined zero-energy gaps for all three cases. Bottom row) Spectra under a single-OBC along the x-direction, with bulk/edge modes colored black/red. All cases have trivial first-order topology: The edge modes of cases A and C do not traverse the gap, and case B does not even have edge modes.  }
\label{edge}
\end{figure*}

Next, we introduce a boundary perpendicular to the x-axis, such that $k_2$ remains a good ``quantum number" (Fig.~\ref{edge} bottom row). While edge modes (red) now appear in cases A and C, they do not traverse the gap. This indicates constant first-order polarization and hence trivial \emph{first-order} $\mathbb{Z}_2$ topology, which is expected from our simple lattice structure devoid of effective pseudospin-orbit coupling~\cite{albert2015topological,ningyuan2015time}.

\begin{figure}
\centering
\includegraphics[width=\linewidth]{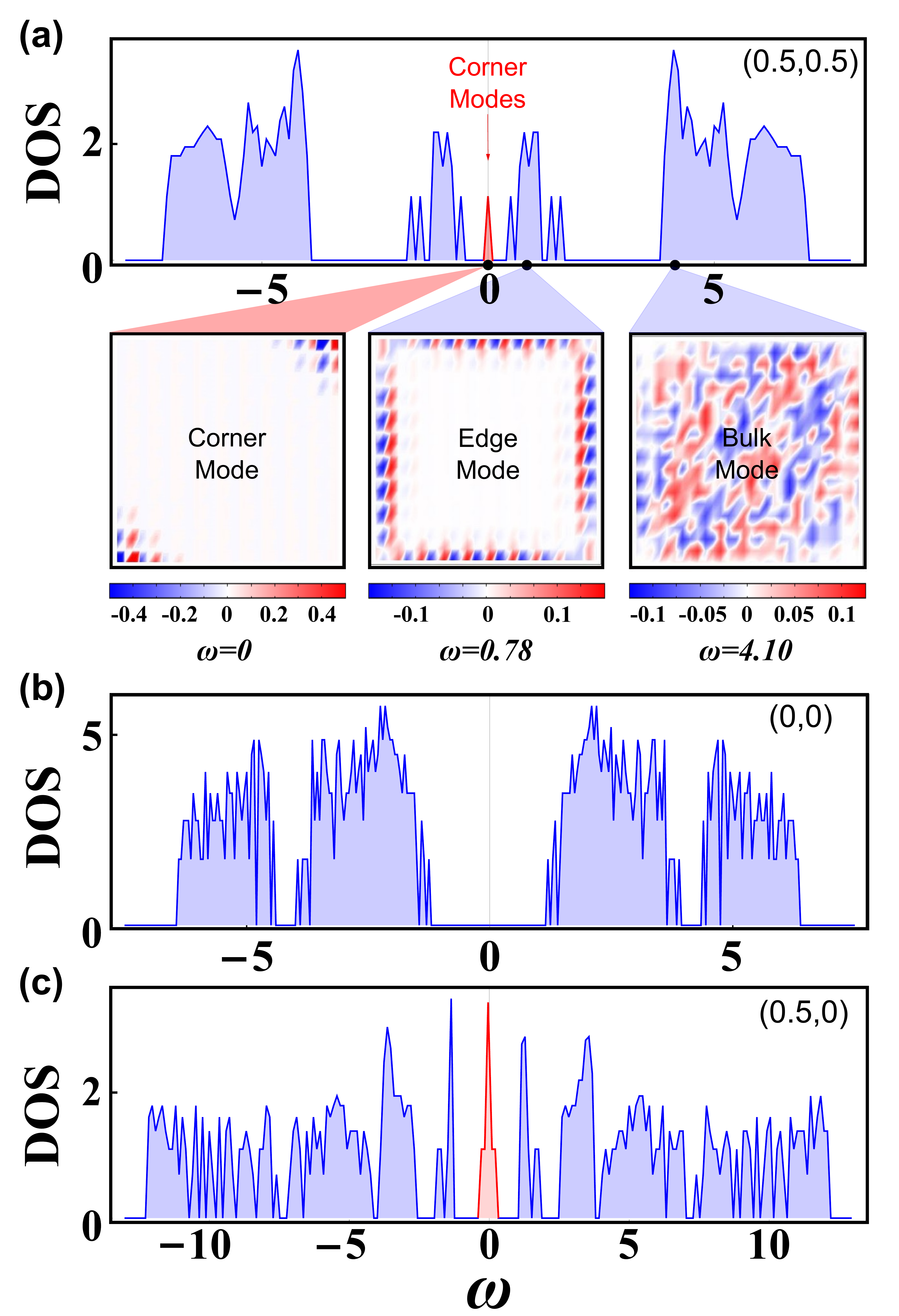}
\caption{Double OBCs density of states (DOS) for cases A to C (a to c) as a function of energy $\omega$, calculated on a finite lattice with $10\times 10$ unit cells. Midgap modes exist within the bulk gap for cases A and C, but only case A has well-separated doubly degenerate corner zero modes. 
Its spatial wavefunction distribution is illustrated at frequencies $\omega=0$ (left) $\omega=0.78$ (middle) and $\omega=4.10$ (right), where it is dominated by corner, edge and bulk modes respectively. The DOS vertical axis is plotted on a logarithmic scale. 
}
\label{dos}
\end{figure}

What is interesting is that, after taking OBCs in both x and y directions (double OBCs), second-order topological corner modes can still appear even though the edge modes with a single OBC do not exhibit topological  polarization. As shown in Fig.~\ref{dos}, such corner modes in cases A and C, but not B.
In case A (Fig.~\ref{dos}(a)), we observe  a two-fold degenerate density of states (DOS) peak at energy $\omega= 0$, each copy corresponding to a corner mode plotted in the lower left panel. Other DOS peaks away from zero energy but within the bulk gap correspond to edge modes. Both corner and edge modes do not exist in case B (Fig.~\ref{dos}(b)), which only exhibit bulk modes. Indeed, without edge modes from single OBCs, corner modes cannot possibly appear when another open boundary is introduced. Case C (Fig.~\ref{dos}(c)) is somewhat similar to case A, but its zero energy modes are not isolated from the other modes, and hence do not form well-defined corner modes. In the following, we shall explain and substantiate these observations through topological arguments. 



\section{Topological characterization of corner modes}

We now briefly recap the theory of higher-order topological polarization before describing a $\mathbb{Z}_2\times\mathbb{Z}_2$ topological classification of our brick lattice corner modes different from that in existing literature.

\subsection{First-order polarization}

First, we introduce the notion of topological (Wannier) polarization. Consider a 2D Hamiltonian $H(k_1,k_2)$ with OBC in the x-direction, such that its eigenstates $|\Psi(k_2)\rangle$ are indexed by $k_2\in [0,2\pi)$, which remains a good quantum number. Of central importance is the projected periodic position operator
\begin{equation}
\hat X_P(k_2)=\hat P(k_2) e^{2\pi i\hat{x}/N_x}\hat P(k_2),
\end{equation}
where $\hat{x}$ is the usual position operator, $\hat P(k_2)=|\Psi(k_2)\rangle\langle \Psi(k_2)|$ is the projection onto a chosen $|\Psi(k_2)\rangle$ band and $N_x$ is the number of unit cells along the x-direction.
The \emph{first-order} polarization $\langle x(k_2)\rangle$ is given by the rescaled phase of the eigenvalues of $\hat X_P(k_2)$:
\begin{equation}
\langle x(k_2) \rangle =\frac{N_x}{2\pi} Im \log{\langle\Psi(k_2)|e^{2\pi i\hat{x}/N_x}|\Psi(k_2)\rangle}
\end{equation}
For a Hamiltonian with $b$ bands, there exists $bN_x$ eigenvalues of $X_P(k_2)$, but only $b$ of them are independent: the rest are translated by a phase of $2\pi/N_x$, and as such correspond to the same polarization~\cite{qi2011generic,lee2015free} $\langle x(k_2) \rangle$.

Physically, the polarization is the center-of-mass position of its corresponding $\hat X_P(k_2)$ eigenstate, which is also a maximally localized Wannier function for any given $k_2$\cite{kivelson1982wannier,marzari1997maximally,yu2011equivalent,asboth2016short}. Hence it is also called the Wannier center. For a Fermi gas of electrons, the polarization tells us, through the Laughlin gauge argument, how charge within the occupied Fermi sea is inevitably topologically ``pumped" by an electric field that translates $k_2$. Numerically, the Wannier centers can be computed via the Wilson loop operator $W_{k_1+2\pi\leftarrow k_1}$, as detailed in Appendix \ref{sec:wilsonloop}.

In our brick lattice with time reversal symmetry, the band topology is characterized by a $\mathbb{Z}_2$ invariant~\cite{kane2005z,yu2011equivalent,fu2007topological} which can be read from the spectral flow of the polarization~\cite{yu2011equivalent}. Specifically, the $\mathbb{Z}_2$ index is trivial/non-trivial depending on whether the $\langle x(k_2) \rangle$ eigenvalues ``switch partners" as $k_2$ varies over half a period, i.e from one time reversal invariant point to the other. This is equivalent to checking whether a particular $\langle x(k_2) \rangle$ Wannier center trajectory crosses an arbitrary line parallel to the $k_2$ axis an even/odd number of times as $k_2$ varies over a period.

In general, the polarization flow bears a one-to-one correspondence with the energy spectral flow: for each pair of Wannier centers that switch partners, there also exist a pair of gapless edge modes that switch partners and traverse the bulk gap. In particular, a gapped OBC spectrum can contain only $\mathbb Z_2=0$ bulk bands, as in all of the cases plotted in the bottom row of Fig.~\ref{edge}. They can possess either edge modes that do not traverse the gap (cases A and C), or no edge modes at all (case B). These behaviors are reflected in their polarization trajectories shown in Fig.~\ref{Wannier}. While none of them exhibit partner switching and are hence all $\mathbb Z_2$ trivial, cases A and C both possess polarizations that fluctuate about $0.5$, indicative of midgap localization tendencies of their respective OBC edge modes.


\begin{figure}
\centering
\includegraphics[width=\linewidth]{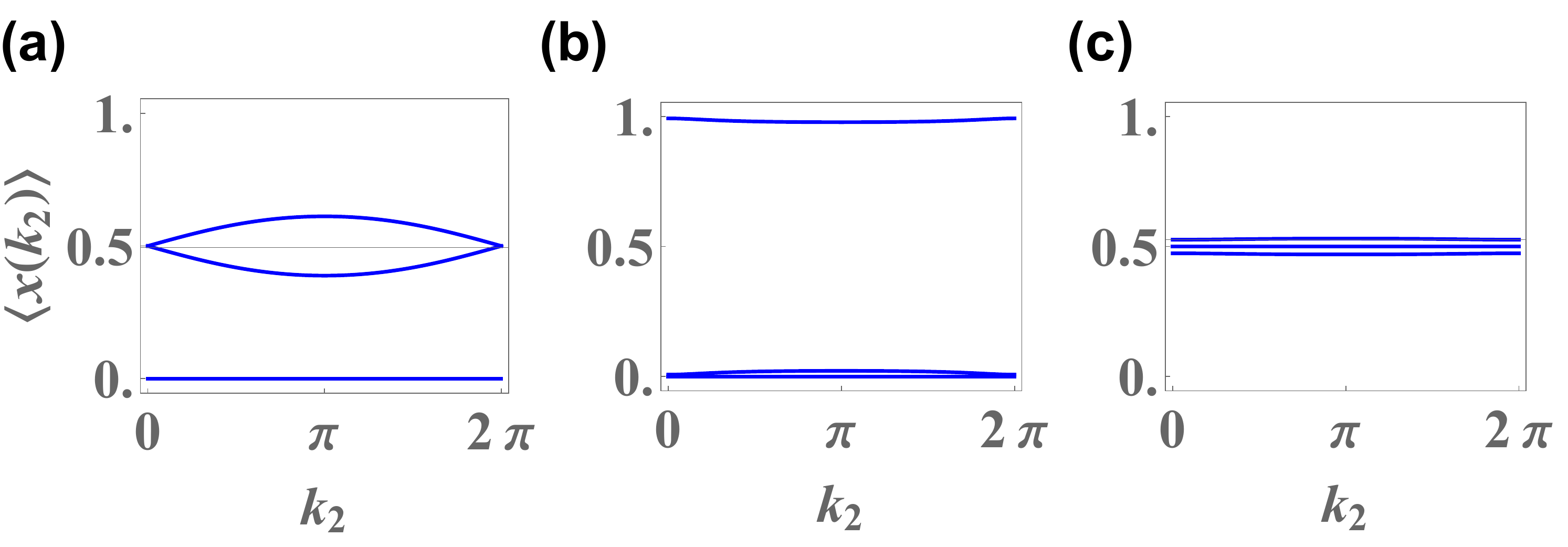}
\caption{Evolution of Wannier centers $\langle x(k_2)\rangle$ for cases A to C (plots a to c respectively) over one period of $k_2$. For all cases, there is no partner switching, and a dispersionless trajectory always exists. Cases A and C, which contain edge modes, also have Wannier centers hovering around $0.5$. These first-order polarizations should not be confused with the second-order polarizations $p_y^j$, which define the second-order $\mathbb Z_2\times \mathbb Z_2$ via Eq.~\ref{Z2Z2}.}
\label{Wannier}
\end{figure}

\subsection{Second-order polarization and $\mathbb{Z}_2\times \mathbb{Z}_2$ classification of corner modes}

To understand how topological corner modes can arise from trivial single OBC $\mathbb Z_2$ edge modes, we now introduce the concept of second-order quadrupole polarization\cite{Benalcazar2017Quantized}. The main idea is to use the gapped (first-order) Wannier bands as the ``bulk" bands of a new effective system, and apply the machinery of Wannier polarization on it to obtain the \emph{second-order} polarization properties of the original system. This procedure can of course be repeated ad infinitum to obtain higher-order polarizations in a higher-dimension system, although we shall stop at the second-order in this work since the brick lattice is 2-dimensional.

More concretely, one divides the set of Wannier centers $\langle x(k_2) \rangle$ into mutually non-intersecting (gapped) sectors, such that intersecting Wannier centers combine to form a single sector~\cite{benalcazar2016quantized,benalcazar2017electric}. Just like gapped bands, each sector is well-separated from the others, and can thus be unambiguously characterized topologically. For each $j$-th Wannier center, $j=1,...,N_F$, where $N_F$ is the number of occupied bands, we can define an effective second-order ``bulk" state $|\omega_{x,\bm{k}}^j\rangle$ in terms of its corresponding Wannier function:
\begin{equation}
|\omega_{x}^j(\bm{k})\rangle=\sum_{n=1}^{N_{F}}|u_{\bm{k}}\rangle[\nu_{x,k_2}^j]^n,
\end{equation}
where $[\nu_{x,k_2}^j]^n$ is the $n$-th component of the $j$-th Wannier function in the basis of occupied bands, and $|u_{\bm{k}}^n\rangle$ is the $n$-th Bloch state. In analogy to the first-order polarization, one can thus compute a \emph{second-order} polarization
\begin{equation}
p_y^j=-\frac{i}{2\pi}\frac{1}{N_x}\sum_{k_1}\log{(\tilde{W}^j_{y,k_1})}
\end{equation}
from the \emph{nested} Wilson loop operator $\tilde{W}^j_{y,k_1}$ formed from $|\omega_{x}^j(\bm{k})\rangle$, as detailed in Appendix \ref{sec:wilsonloop}. The $y$ subscript in $p^j_y$ indicates that it refers to the $y$-direction nested polarization of x-OBC Wannier functions; $p^j_x$ does not necessarily equal $p^j_y$ unless a mirror symmetry maps one boundary to the other.

\begin{figure}
\centering\includegraphics[width=\linewidth]{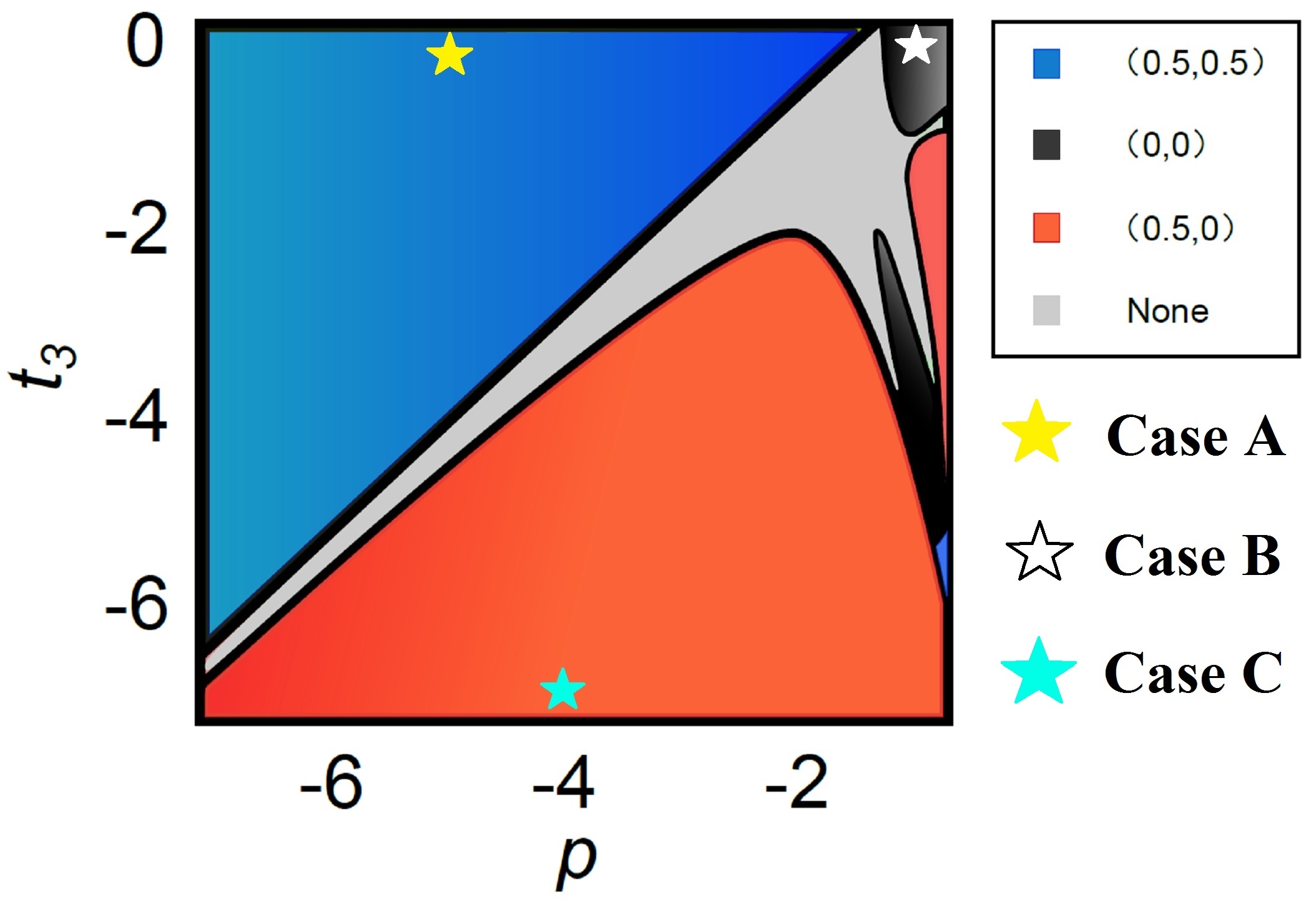}
\caption{Topological phase diagram for our brick lattice, with cases A to C each corresponding to a different $(\mu,\nu)\in \mathbb Z_2\times Z_2$ class. Interestingly, the trivial $(0,0)$ phase occupies a relatively small region, and is separated from the two distinct topological phases via a somewhat larger gapless regime (gray), where the topology is not well-defined. Generally, $|p|>|t_3|$ favors the blue $(0.5,0.5)$ nontrivial phase (and vice versa for $|p|<|t_3|$), illustrating the competition between width and edge couplings across the unit cells.
}
\label{phase}
\end{figure}

The topological class of a second-order (double OBC) system is given by the set of $\mathbb Z_2$ numbers associated with the Wannier sectors. In the well-studied square corner mode lattice~\cite{Benalcazar2017Quantized,benalcazar2017electric} with $N_F=2$ gapped occupied edge bands at half filling, there are two Wannier sectors, and a $\mathbb Z_2\times \mathbb Z_2$ classification can be defined~\footnote{Alternatively, this topological classification can also be obtained through the mirror Chern number~\cite{imhof2018topolectrical}, without reference to higher-order polarization.}. But in our model with $N_F=3$ occupied bands at half filling, a \emph{different}~\cite{Benalcazar2017Quantized} $\mathbb Z_2\times \mathbb Z_2$ classification must be defined. Since there is already a dispersionless Wannier center due to odd $N_F$ and $M_x$ symmetry, we shall let it be in its own Wannier sector with corresponding second-order polarization $p^1_y$ (a flat trajectory in each plot of Fig.~\ref{Wannier}). The other two Wannier bands may generically intersect, and shall be taken to form the other sector. Hence we define, for our brick lattice, a new $\mathbb Z_2\times \mathbb{Z}_2$ topological index
\begin{equation}
(\mu,\nu)=(p_y^1,p_y^2+p_y^3)
\label{Z2Z2}
\end{equation}
A phase diagram for the brick lattice is shown in Fig.~\ref{phase} for fixed intra-unit cell couplings $a=t_1=-1$ and variable inter-unit cell couplings $p$ and $t_3$. Case A is deep within the $(\mu,\nu)=(0.5,0.5)$ region with $p_y^1=0.5$ and $p_y^2+p_y^3=0.5$, and host two distinct degenerate corner modes. Case B, which essentially consists of islands dominated by intra-unit cell couplings $a,t$ , is non-topological as expected, with neither edge (Fig.~\ref{edge}) nor corner modes. Case C belongs to the more enigmatic $(0.5,0)$ phase, which is encouraged by a dominant $t_3$. To gain some intuition, consider the extreme limit of large $|t_3|$ and small $|p|$, where the brick lattice essentially splits into weakly coupled 1D Su-Schrieffer-Heeger (SSH) ladders with strong/weak couplings $t_3$ and $t_1$, and relatively weak ``rungs" composed of two successive $a$ couplings (Fig.~\ref{lattice}). In this quasi-1D limit, corner modes obviously should not exist, although a continuum of boundary modes at the ends of each ladder still gives rise to $1/2$ polarization. In this sense, the $(0.5,0)$ phase can be regarded as the ``horizontal half" of the $(0.5,0.5)$ phase, although the above analogy quickly becomes inaccurate away from the extreme limit. Finally, we note that the various topological phases are usually not adjacent to each other: to transform from one topological phase to another, the requisite bandgap closure may last indefinitely long, i.e. if the parameters are transformed along the gray strip $t_3\approx p$.

\section{Corner modes from effective 1D edge picture}
To more intuitively understand the origin of the corner modes, we now consider cases where the corner mode can be largely explained with a 1D edge picture. Instead of invoking the rather abstract nested polarizations, we attempt to visualize corner modes as the intersections of the boundary modes of 1D SSH-like edges.

\begin{figure}
 \centering
 \includegraphics[width=\linewidth]{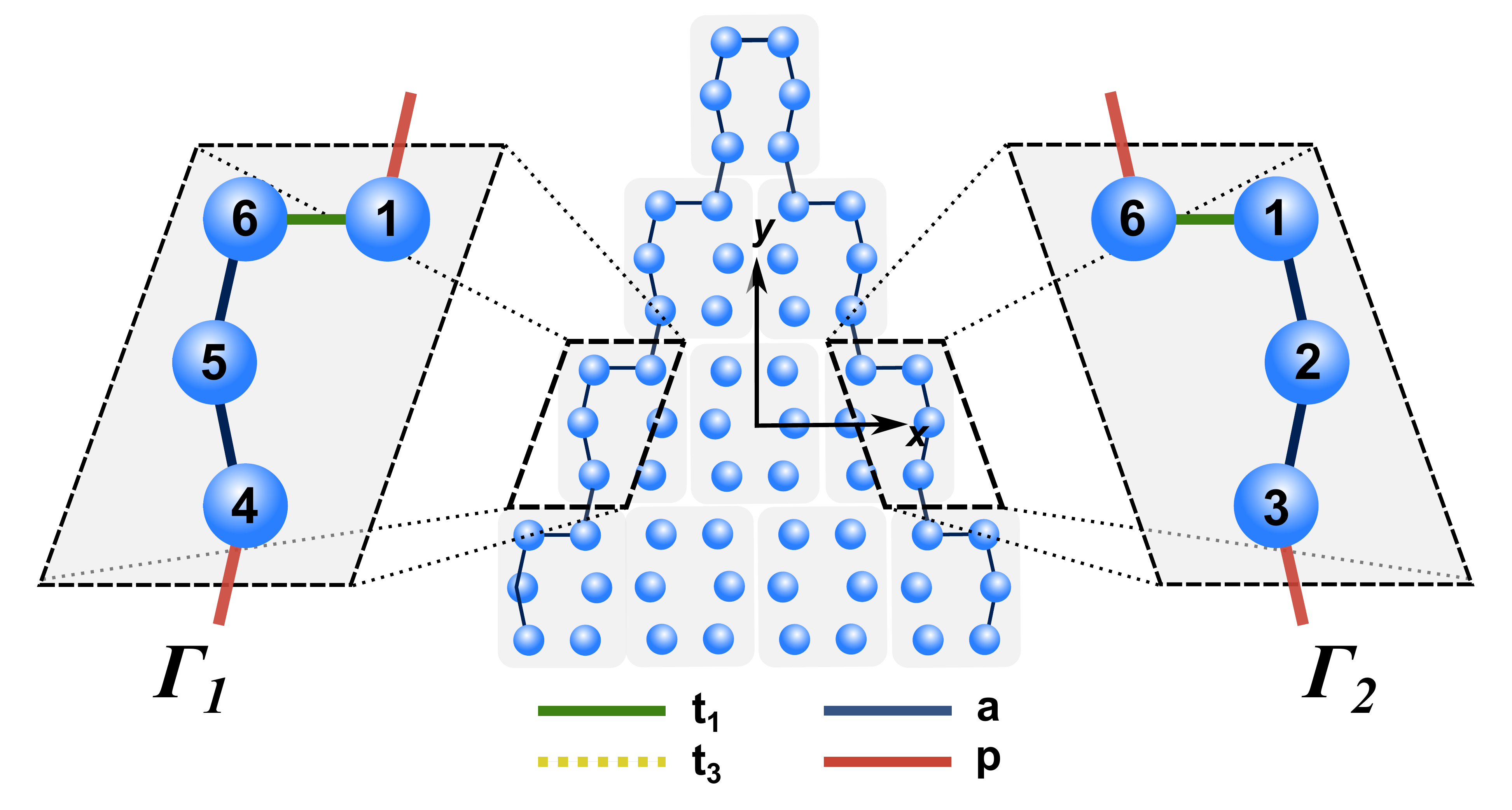}
 \caption{One-dimensional armchair chains $\Gamma_1$ and $\Gamma_2$ at the edges of both boundaries, which are equivalent due to lattice mirror symmetry. The 4 inequivalent "atoms" in their unit cells couple according to $H_\text{edge}$ (Eq.~\ref{1dHamiltonian}), whose gap closure at $p=t_1$ for $t_3=0,a=-1$ agrees with the  phase diagram for the \emph{full} brick lattice (Fig.~\ref{phase}).  }
 \label{edgelattice}
  \end{figure}

 \begin{figure}
 \centering
 \includegraphics[width=\linewidth]{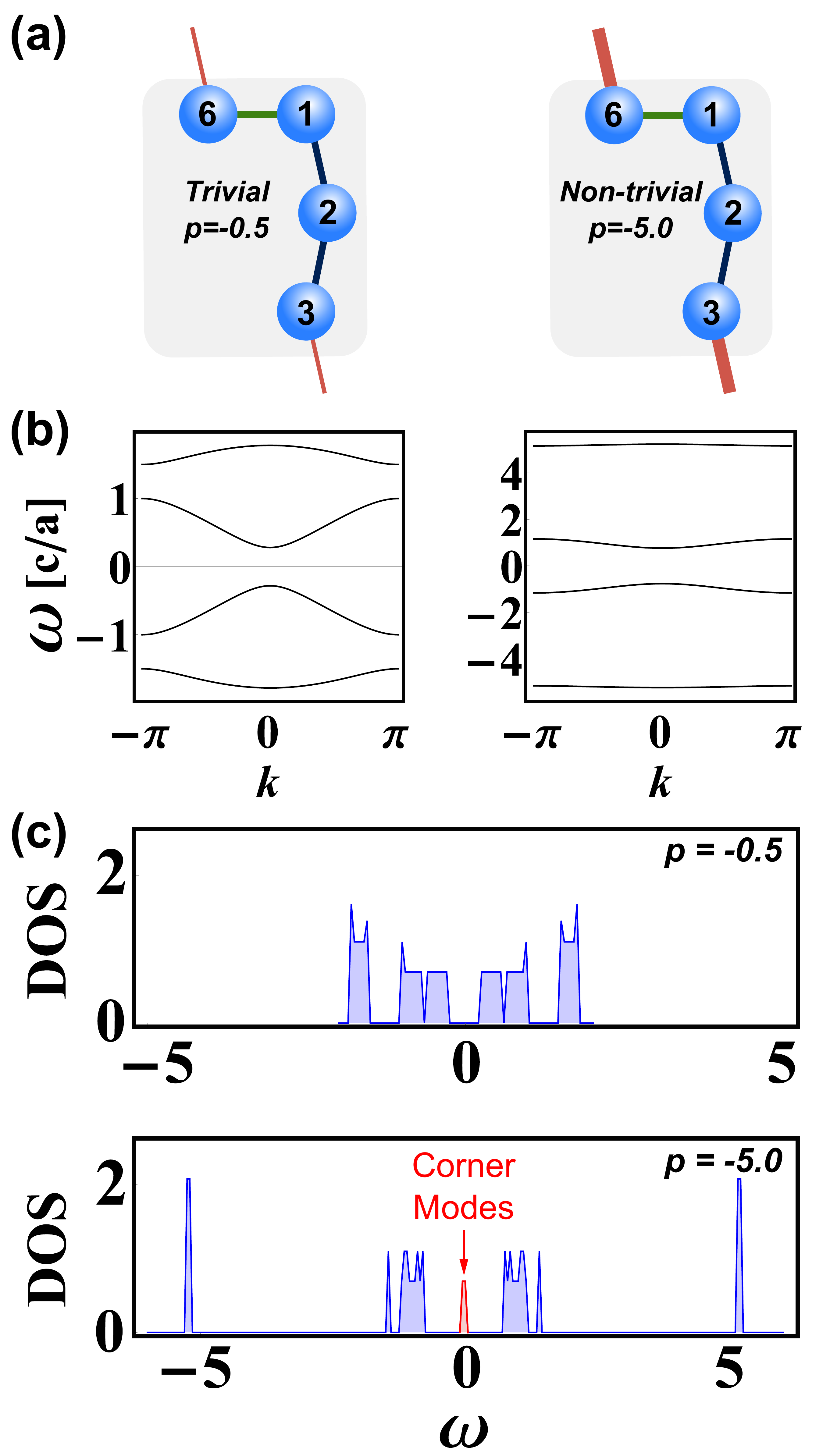}
 \caption{a)Illustration of one-dimensional armchair chains at the edges (we take $\Gamma_2$ for example), with different coupling parameters $p$. (b) The bulk modes and c) DOS plots reveal the existence of corner modes in the one-dimensional armchair chains with topologically non-trivial $p$, with qualitative agreement with the DOS of the full brick lattice in Fig.~\ref{dos}. }
 \label{1d}
  \end{figure}
The double OBCs in our brick lattice produces armchair-like edges in both directions,
as shown in Fig.\ref{edgelattice}. Evidently, the edgemost couplings form SSH-like chains along each edge, each with four sites per unit cell:
\begin{equation}
 H_\text{edge}(k)=\left(
 \begin{array}{cccc}
 0 & t_1 & 0 &p\, e^{-ik}\\
 t_1 & 0 & a &0\\
 0&a&0&a\\
 p \,e^{ik} &0&a&0
 \end{array}\right)
\label{1dHamiltonian}
\end{equation}
In the chains $\Gamma_1$ and $\Gamma_2$ as shown, the basis in $H_\text{edge}$ are taken to be sites $1,6,5,4$ and $6,1,2,3$ respectively.

Like the well-known SSH model, this 4-band model contains topological zero modes when the inter-unit cell coupling $p$ is larger than the intra-unit cell coupling $t_1$. This can be seen from the analytic expression of its eigenenergies $\omega^2=a^2+Q^2\pm\sqrt{a^4+Q^4+2a^2pt_1\cos k}$ with $Q^2=(p^2+t_1^2)/2$, which gives the \emph{only} gap closure and hence possible topological phase transition at $t=p_1$. In other words, the $a$ intra-unit cell couplings are ``spectators" that play no part in determining the topology, and leave behind an SSH-like dimerization mechanism for topological boundary modes. Setting $t_1=a=-1$ as before, we see that the topological phase transition point $p=a=-1$ for $H_\text{edge}$ indeed agrees with the phase diagram of the full brick lattice in Fig.~\ref{phase}. Indeed, as shown in Fig.~\ref{1d}c, its DOS also agrees qualitatively with that of the full brick lattice in Fig.~\ref{dos}, with corner modes comprising superposed SSH-like boundary modes from both chains $\Gamma_1$ and $\Gamma_2$. Note that his admittedly rudimentary edge model completely ignores the effects of coupling between adjacent chains, and thus cannot predict the effects of $t_3$. A more detailed analysis with these neighboring couplings may provide intuition for the entire phase diagram, as has been done for the square corner mode model~\cite{li2018direct}.


\section{Effect of breaking nonsymmorphic symmetry}

As previously emphasized, a hallmark of our brick lattice is its nonsymmorphic symmetry in addition to its two commuting mirror symmetries. Below, we show that with our lattice structure, the nonsymmorphic symmetry $g_y=\{m_y|\tau_x\}$ is \emph{essential} in protecting the corner zero modes, unlike the extensively studied square corner mode lattice~\cite{Benalcazar2017Quantized} which requires only the two mirrors symmetries $M_x$ and $M_y$.

As illustrated in Fig.~\ref{nonsym}a, we break the nonsymmorphic symmetry $g_y=\{m_y|\tau_x\}$ by removing the $t_1$ couplings of alternate original unit cells (green $\rightarrow$ white) i.e. sites $1,6$ and $3,4$ are no longer coupled by $t_1$. Doing so, the mirror symmetries $M_x$ and $M_y$ are obviously preserved, since the $t_1$'s are removed symmetrically within each unit cell. However, nonsymmorphic symmetry is broken because site $12$ no longer maps identically to site $4$, and ditto for site $11$ to site $5$, etc. From Fig.~\ref{nonsym}b,c, we no longer observe well-defined zero modes in the DOS. This destruction of the corner zero modes is expected from the previous effective edge picture, which gives two inequivalent SSH-like chains that do not ``dimerize" in the same way.

\begin{figure}
\centering\includegraphics[width=\linewidth]{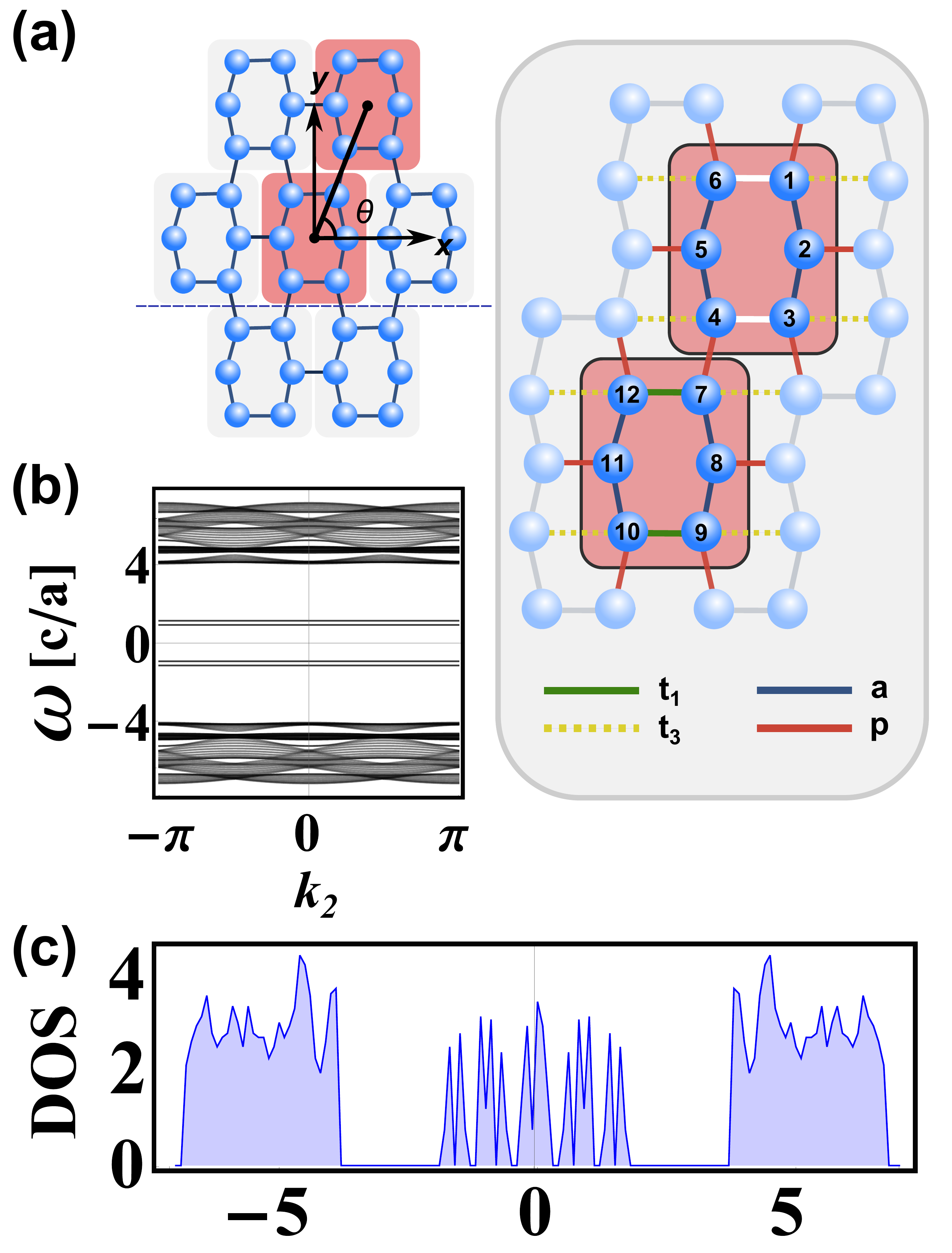}
\caption{(a) Illustration of a modification to the brick lattice that breaks the nonsymmorphic symmetry $g_y$ but preserves the mirror symmetries $M_x$ and $M_y$. Each unit cell now consists of 12 sites, with half of the $t_1$ (green) couplings removed. b) Single OBC spectrum and c) double OBCs DOS, showing the absence of isolated zero energy topological modes. }
\label{nonsym}
\end{figure}

\section{Experimental proposal via RLC circuits}

Finally, we briefly discuss how to experimentally realize a brick lattice and measure its corner modes. Of various possible platforms in photonic, mechanical and acoustic systems\cite{zhang2017entangled,lee2018topological,Serra2018Observation}, an RLC circuit realization is arguably the least challenging, with experimental smoking gun being easily performed impedance experiments~\cite{ningyuan2015time,helbig2018band,wang2018topologically,lu2018probing,imhof2018topolectrical}. Since this approach is already quite mature with a similar corner mode circuit experiment performed last year~\cite{imhof2018topolectrical}, we shall refer the reader to various excellent references for most of the details~\cite{albert2015topological,lee2018topolectrical,hofmann2018chiral}.

In a circuit, the physics are most directly described via Kirchhoff's law, which can be put into a matrix form
\begin{equation}
I_a(\omega)=\sum_{b}J_{ab}(\omega)V_b(\omega)
\end{equation}
where $I_a(\omega)$ and $V_b(\omega)$ are the frequency-space net input current and electrical potential at nodes $a,b$ respectively. $J_{ab}(\omega)$ is the circuit Laplacian that captures the circuit behavior. For our purpose, $J_{ab}(\omega)$ will replace the Hamiltonian, such that the DOS and energy spectrum now refers to that of the Laplacian.

To realize our brick lattice (Eq.~\ref{Hamiltonian}) with a Laplacian, one simply substitutes each coupling by a capacitor proportional to its value, such that a coupling $H_{jk}=-p$, $p>0$ becomes the admittance contributions $J_{jj}(\omega)=J_{kk}(\omega)=-i\omega p$ and $J_{jk}(\omega)=J_{kj}(\omega)=i\omega p$, $\omega$ the AC frequency. To independently control the onsite couplings, we also connect grounded inductors $L_e$ or $L_f$ to each site, such that they acquire onsite admittance contributions of $(i\omega L_e)^{-1}$ or $(i\omega L_f)^{-1}$. Made out of capacitors of capacitances $a,p,t_1,t_3$ and grounding inductors described below, the brick circuit possesses a Laplacian of the form
\begin{widetext}
\begin{equation}
J(k_1,k_2;\omega)=-i\omega\left(
\begin{array}{cccccc}
      e' & a & 0 & p\, e^{i k_1}& 0 & t_1+t_3 e^{i(k_1+k_2)}\\
      a & f' & a & 0 & p\, e^{i(k_1+k_2)} &0\\
     0 & a & e' & t_1+t_3 e^{i(k_1+k_2)} & 0 & p\, e^{ik_2}\\
     p\, e^{-ik_1} & 0 & t_1+t_3 e^{-i(k_1+k_2)} & e' & a & 0\\
     0 & p\,  e^{-i(k_1+k_2)} & 0 & a & f' & a\\
     t_1+t_3 e^{-i(k1+k2)} & 0 & p\, e^{-ik_2}& 0 & a & e'
\end{array}
\right),
\label{Laplacian}
\end{equation}
\end{widetext}
with $e'=\frac1{\omega^2L_e}-a-p-t_1-t_3$ and $f'=\frac1{\omega^2L_f}-2a-p$. By tweaking $L_e$ and $L_f$, one can easily make them equal, such that the onsite admittances become a constant shift of the Laplacian eigenvalues, analogous to the chemical potential.

To detect the corner modes, one measures the impedance~\cite{lee2018topolectrical}
\begin{equation}
\begin{aligned}
Z_{ab}(\omega)&=\frac{V_a-V_b}{I_0}\\
&=\sum_n \frac{|\phi_n(a)-\phi_n(b)|^2}{j_n(\omega)}
\end{aligned}
\end{equation}
between two nodes $a,b$ with respect to a current $I_0$ entering from $a$ and leaving from $b$. The second line is defined via $J_{ab}(\omega)=\sum_n j_n(\omega)|\phi_n(a)\rangle\langle\phi_n(b)|$, the expansion of the Laplacian into its eigenmodes.  Most salient from this key expression is that zero modes $j_n(\omega)\approx 0$ give rise to large divergences, which are also known as topolectrical resonances. By measuring the impedance between two points near a corner, corner zero modes can be easily identified as large impedances/resonances.

\section{Conclusion}
Compared to well-known higher-order lattices like the square corner mode lattice, our brick lattice is fundamentally different in two ways: its corner zero modes requires nonsymmorphic symmetry in addition to two mirror symmetries, and it has an odd number of occupied bands that necessitates a new definition of the $\mathbb Z_2\times \mathbb Z_2$ topological invariant. In addition to trivial gapped and gapless phases, we also uncovered two distinct topological phases: $(\mu,\nu)=(0.5,0.5)$ with distinct corner modes, and $(\mu,\nu)=(0.5,0)$ hosting continuum boundary modes and adiabatically connected to weakly coupled SSH ladders. We conclude our work by describing how brick lattice corner modes can be realized and easily detected in a circuit setup, a platform that has proved to be experimentally accessible and amenable to interesting non-linear, non-Hermitian or Floquet generalizations~\cite{wang2018topologically,lee2018hybrid,ezawa2018electric,ezawa2018non}. 

\begin{acknowledgments}
Yuhan Liu and Yuzhu Wang contributed equally to this work. Ching Hua Lee thanks Linhu Li for discussions.
\end{acknowledgments}

\bibliography{ref}

\newpage
\appendix
\begin{widetext}

\section{The Wilson loop}
\subsection{ Wilson loop over occupied energy bands\label{sec:wilsonloop}}

In the main text, we have alluded to using the Wilson loop to compute the Wannier center evolution of a given Hamiltonian. Here we show a detailed description of the procedure, mainly following~\cite{asboth2016short,yu2011equivalent} and the supplement of~\cite{Benalcazar2017Quantized}. Fixing $k_2$ such that the system is effectively one-dimensional, the projection operator $\hat{P}(k_2)$ to the occupied bands (from $n=1$ to $n=N_F$) is
\begin{equation}
\hat{P}(k_2)=\sum_{k_1} \sum_{n=1}^{N_F}\left|\Psi_n(k)\rangle\langle \Psi_n(k)\right|=\sum_{k_1}\left|k\rangle\langle k\right|\otimes \hat{P}(k)
\end{equation}
where $\hat{P}(k)=\sum_{n=1}^{N_F}\left|u_n(k)\rangle\langle u_n(k)\right|$. We next write down the unitary periodic position operator of the occupied bands is defined as,
\begin{equation}
\hat{X_P}(k_2)=\hat{P}(k_2)\hat X\hat{P}(k_2)
\label{PXP}
\end{equation}
where $\delta_{k}=2\pi/N_x$ and $\hat{X}=e^{i\delta_{k_1}\hat{x}}$.
Using $\langle \Psi_{n'}(k')|\hat{X}|\Psi_n(k)\rangle=\delta_{k+\delta_{k},k'}\langle u_{n'}(k+\delta_{k})|u_n(k)\rangle$, and substituting the above definition of $\hat{P}(k_2)$, we get,
\begin{equation}
\hat{X}_P(k_2)=\sum_{k_1} \sum_{n',n=1}^{N_F}\langle u_{n'}(k+\delta_k)|u_n(k)\rangle \cdot |\Psi_{n'}(k+\delta k)\rangle\langle \Psi_n(k)|
\end{equation}
The summation over k has $N_x$ terms, so the above operator can be expressed as a $N_F\times N_x$ matrix. If we define matrix $G_k$ with component $[G_k]^{mn}=\langle u_n(k+\delta k)|u_m(k)$, it is not unitary because $N_x$ is finite. To facilitate the numerical computation, we can do the singular value decomposition $G=UDV^{\dagger}$  where $D$ is a diagonal matrix. If we define $F_k=UV^{\dagger}$, we get a unitary matrix which equals to $G_k$ in the thermodynamic limit, and we can write the operator $\hat{X}_P$ in the thermodynamic limit case, under the $N_x\times N_F$ basis of $|\Psi_n(k)\rangle$:
\begin{equation}
\hat{X}_P(k_2)=\left(
\begin{array}{ccccc}
0 &0 &0 &\cdots & F_{k_N} \\
F_{k_{i}} &0&0&\cdots &0\\
0 & F_{k_{ii}} &0 &\cdots &0\\
\vdots & \vdots & \vdots & \ddots & \vdots\\
0 &0&0&\cdots &0
\end{array}
\right),
\end{equation}
where $k_{i}=0,k_{ii}=\delta k,\cdots,k_N=\delta k(N_x-1)$. Each matrix $F$ is a $N_F\times N_F$ matrix. We write its eigenvector in terms of a $1\times N_F$ block, namely,
\begin{equation}
\hat{X}_P\left(
\begin{array}{c}
\nu_{k_{i}}\\
\nu_{k_{ii}}\\
\nu_{k_{iii}}\\
\vdots\\
\nu_{k_N}
\end{array}
\right)^j=E^j
\left(
\begin{array}{c}
\nu_{k_{i}}\\
\nu_{k_{ii}}\\
\nu_{k_{iii}}\\
\vdots\\
\nu_{k_N}
\end{array}
\right)^j
\end{equation}
The Wilson loop operator is defined as
\begin{equation}
W_{k+2\pi\leftarrow k}=F_{k+2\pi-\delta k}F_{k+2\pi-2\delta k}\cdots F_{k+\delta k}F_{k}
\end{equation}
By recursively applying the above equations to the eigenvector, we can derive the eigenvalue equation
\begin{equation}
W_{k_1+2\pi\leftarrow k_1}|\nu_{x,\bm{k}}^j\rangle=(E^j)^{N_x}|\nu_{x,\bm{k}}^j\rangle
\label{Wilson}
\end{equation}
Here we write $|\nu_{\bm{k}}^j\rangle$ as $|\nu_{x,\bm{k}}^j\rangle$ to denote that the Wilson loop is taken along $\hat{x}$. It should be noticed that although the eigenstates $|\nu_{\bm{k}}^j\rangle$ are different for different $k$, their eigenvalues are the same for a fixed $k_2$. So if we only care about the eigenvalue, we can choose any $k_1$ the starting point of the Wilson loop. If we have $N_F$ occupied bands, we can solve Eq.(\ref{Wilson}) to get $N_F$ different $E^N$. Looking back on the definition of $\hat{X}_P$ in Eq. (\ref{PXP}), we can relate the phase of $(E^j)^{N_x}$ to $\langle x \rangle$ as in the main text.

Fig. \ref{Wannier} of the main text plots the phase of $(E^j)^{N_x}$ of different $k_2$. Since the Hamiltonian possesses pseudo time reversal symmetry, we only need to plot $k_2$ from 0 to $\pi$, with the part from $-\pi$ to 0 related by symmetry.

\subsection{ Nested Wilson loop over Wannier sectors}
We define the Wannier basis

\begin{equation}
|\omega_{x}^j(\bm{k})\rangle=\sum_{n=1}^{N_{F}}|u_{\bm{k}}^n\rangle[\nu_{x,\bm{k}}^j]^n
\end{equation}
as in the main text, and use it to calculate the nested Wilson loop $\tilde{W}_{y,k_1}$ in a similar way as the (first-order) Wilson loop:

\begin{equation}
\tilde{W}^j_{y,k_1}=\tilde{W}^j_{k_2+2\pi\leftarrow k_2}=\tilde{F}^j_{k_2+2\pi-\delta k}\tilde{F}^j_{k_2+2\pi-2\delta k}\cdots \tilde{F}^j_{k_2+\delta k}\tilde{F}^j_{k_2}
\label{Wy}
\end{equation}
where $[\tilde{F}^j_{k_2}]^{mn}=\langle \omega^{j,m}_x(k_1,k_2+\delta k)|\omega^{j,n}_x(k_1,k_2)\rangle$, which is independent of $x$.
\end{widetext}

\end{document}